\documentclass[prd,twocolumn,floatfix,amsmath,nofootinbib,amssymb,floatfix]{revtex4}
\usepackage{graphicx,color,dcolumn,booktabs,bm}
\usepackage{longtable,lscape}
\usepackage{pdfpages}
\usepackage{txfonts}
\usepackage{overpic}
\usepackage{amssymb}
\usepackage{makecell}
\usepackage{indentfirst}
\usepackage{feynmf}   
\usepackage{slashed}  
\usepackage{cases}
\usepackage{color}
\usepackage{multirow}
\usepackage{makecell}
\usepackage{threeparttable}
\usepackage{epstopdf}
\usepackage{enumerate}
\usepackage{diagbox}
\usepackage{graphicx,color,dcolumn,booktabs,bm}
\usepackage[colorlinks,
            citecolor=blue,
            anchorcolor=red,
            menucolor=red,
            linkcolor=red,
            filecolor=red,
            runcolor=red,
            urlcolor=blue,
            frenchlinks=red]{hyperref}
\usepackage{tikz}

\begin{document}

\title{Roles of $f_{0}(500)$ and $f_{0}(980)$ in the $D_{(s)}^{+}\rightarrow\pi^{+}\pi^{+}\pi^{-}$ decays}

\author{Zhong-Yu Wang$^{1}$}
\email{zhongyuwang@gzu.edu.cn}

\author{Wei Liang$^{2}$}
\email{212201014@csu.edu.cn}

\affiliation{
$^1$College of Physics, Guizhou University, Guiyang 550025, China \\
$^2$School of Physics, Central South University, Changsha 410083, China \\
}

\date{\today}

\begin{abstract}

With the recent measurements of the $D_{s}^{+}\rightarrow \pi^{+}\pi^{+}\pi^{-}$ and $D^{+}\rightarrow \pi^{+}\pi^{+}\pi^{-}$ decays by the LHCb Collaboration, we study these two decay processes by considering the final state interaction formalism.
Taking into account the external and internal $W$-emission dominant mechanisms at the quark level, our model can naturally explain why the $D_{s}^{+}$ decay only has a contribution from the $f_{0}(980)$ resonance, while the $D^{+}$ decay has contributions from both the $f_{0}(500)$ and $f_{0}(980)$ states.
The magnitudes and phases of $\pi^{+}\pi^{-}$ for $S$-wave amplitudes in our model are in agreement with the experimental measurements.
These results support the interpretations of $f_{0}(500)$ and $f_{0}(980)$ as the $\pi\pi$ and $K\bar{K}$ resonances, respectively, where they are dynamically generated from the $S$-wave pseudoscalar-pseudoscalar meson interactions within the chiral unitary approach. 

\end{abstract}
\maketitle

\section{Introduction}\label{sec:Introduction}

Since the discovery of $X(3872)$ by the Belle Collaboration in 2003 \cite{Belle:2003nnu}, a number of new hadronic states have been observed by various high-energy physics experiment groups, as summarized in the reviews \cite{Liu:2013waa,Hosaka:2016pey,Chen:2016qju,Richard:2016eis,Lebed:2016hpi,Olsen:2017bmm,Guo:2017jvc,Liu:2019zoy,Brambilla:2019esw,Meng:2022ozq,Chen:2022asf,Liu:2024uxn}.
Understanding the structure and component of new hadronic states is an important and fascinating research topic that has attracted considerable attention from the whole community.
In particular, the light scalar mesons are more interesting and complex because their masses are in a narrow range.
Some states have many different explanations, such as conventional mesons with quark content $q\bar{q}$, compact tetraquarks, molecular states, the mixing of different components and so on.
Fortunately, the decays of charm and beauty hadrons provide a valuable opportunity to understand the light scalar mesons.

For example, the accumulation of experimental data on the decays of $D$ and $D_{s}$ provides an important platform for theoretical research.
In recent years, the LHCb and BESIII Collaborations have observed the contributions of the $f_{0}(500)$, $f_{0}(980)$, $a_{0}(980)$, and $K_{0}^{*}(700)$ states in the processes of $D^{+}$ decaying into $K^{-}K^{+}K^{+}$, $\pi^{+}\pi^{0}\eta$, $\pi^{+}\eta\eta$, $D^{0}$ decaying into $\pi^{+}\pi^{-}\eta$, $\pi^{0}\eta\eta$, and $D_{s}^{+}$ decaying into $\pi^{+}K^{+}K^{-}$, $\pi^{+}\pi^{-}K^{+}$, $\pi^{+}\pi^{0}\eta$, etc. \cite{LHCb:2019tdw,BESIII:2019xhl,BESIII:2024tpv,BESIII:2018hui,BESIII:2020ctr,BESIII:2022vaf,BESIII:2019jjr}, which provide data support for studying the properties of these earlier discovered scalar states.
The two-body invariant mass distributions and some branching fractions measured by experiments can be reproduced by the the chiral unitary approach (ChUA), where these scalar mesons are generated from the $S$-wave interactions of the pseudoscalar and pseudoscalar mesons \cite{Roca:2020lyi,Duan:2020vye,Ikeno:2021kzf,Wang:2021kka,Wang:2021naf,Wang:2021ews,Dai:2023jix,Liang:2023ekj,Molina:2019udw,Ling:2021qzl,Bayar:2023azy}.
In contrast, Refs \cite{Achasov:2017edm,Cheng:2022vbw,Achasov:2024nrh,Cheng:2024zul} explain these experimental data under the picture of the four-quark model.

In addition, an enhancement near $1.7$ GeV in the $K_{S}^{0}K_{S}^{0}$ invariant mass spectrum in the $D_{s}^{+}\rightarrow K_{S}^{0}K_{S}^{0}\pi^{+}$ decay has been observed by the BESIII Collaboration \cite{BESIII:2021anf}. 
This finding implies the existence of the $a_{0}(1710)$ state, since the quantum numbers of $f_{0}(1710)$ and $a_{0}(1710)$ are the same, the experiment did not distinguish them and labelled them together as $S(1710)$.
Subsequently, similar structure was also observed in the $D_{s}^{+}\rightarrow K_{S}^{0}K^{+}\pi^{0}$ decay \cite{BESIII:2022npc}.
The $a_{0}(1710)$ resonance was confirmed to exist in the $K_{S}^{0}K^{+}$ invariant mass spectrum, which has been called as the $a_{0}(1817)$ state in the existing literatures \cite{BESIII:2022npc,Guo:2022xqu}.
Since the masses of $f_{0}(1710)$ and $a_{0}(1710)$ are around the $K^{*}\bar{K}^{*}$ channel threshold, Refs. \cite{Dai:2021owu,Zhu:2022wzk,Peng:2024ive,Zhu:2022guw,Oset:2023hyt,Wang:2023aza} explain them as the quasi-bound states of $K^{*}\bar{K}^{*}$ with isospin $I=0$ and $I=1$, respectively.
The $K_{S}^{0}K_{S}^{0}$ and $K_{S}^{0}K^{+}$ invariant mass spectra were well reproduced with the final state interaction under the ChUA \cite{Zhu:2022wzk,Peng:2024ive,Zhu:2022guw,Wang:2023aza}.
However, there are still other points of views, such as the conventional meson with a quark content of $q\bar{q}$ \cite{Guo:2022xqu}, the four-quark state \cite{Achasov:2023izs}, and a mixture of two-quark state and tetraquark \cite{Kim:2024atg}. 
The overlapping energy regions of the different structures lead to the greater complexity of the light scalar mesons.

Last year, the LHCb Collaboration reported measurements of the amplitude in the process of $D^{+}\rightarrow\pi^{+}\pi^{+}\pi^{-}$ \cite{LHCb:2022lja}, and the signals of the $f_{0}(500)$ and $f_{0}(980)$ states appear in the magnitude and phase of the $\pi^{+}\pi^{-}$ $S$-wave amplitude.
However, in the $D_{s}^{+}\rightarrow\pi^{+}\pi^{+}\pi^{-}$ decay, only the $f_{0}(980)$ but no $f_{0}(500)$ state has been observed by the BABAR, BESIII, and LHCb Collaborations in Refs. \cite{BaBar:2008nlp,BESIII:2021jnf,LHCb:2022pjv}.
These measurements provide the latest experimental data for us to investigate the nature of the $f_{0}(500)$ and $f_{0}(980)$ resonances.
To describe the amplitudes and phases of the $D^{+}$ and $D_{s}^{+}$ decays, Ref. \cite{Achasov:2022hbh} adopts a phenomenological model to produce the light scalar mesons $f_{0}(500)$ and $f_{0}(980)$ in $\pi\pi$ and $K\bar{K}$ interactions in the final state, which is consistent with the four-quark hypothesis.
In fact, the decay of $D_{s}^{+}\rightarrow\pi^{+}\pi^{+}\pi^{-}$ was studied earlier in Ref. \cite{Dias:2016gou}, where the $f_{0}(980)$ resonance was interpreted as a molecular state.
Although these light scalar mesons have been discovered for many years, the debate about their inside structures and properties are still ongoing.

Inspired by the above experiments, in the present work, we investigate the nature of the $f_{0}(500)$ and $f_{0}(980)$ states in the $D_{(s)}^{+}\rightarrow\pi^{+}\pi^{+}\pi^{-}$ decays.
Starting from the weak decays of $D^{+}$ and $D_{s}^{+}$ at the quark level, we consider the final state interaction to calculate the amplitudes and phases in the $D^{+}\rightarrow\pi^{+}\pi^{+}\pi^{-}$ and $D_{s}^{+}\rightarrow\pi^{+}\pi^{+}\pi^{-}$ decays, where the scalars $f_{0}(500)$ and $f_{0}(980)$ are generated from the $S$-wave pseudoscalar-pseudoscalar meson interaction within the ChUA. 

This paper is organized as follows.
The theoretical formalism of the decays $D^{+}\rightarrow\pi^{+}\pi^{+}\pi^{-}$ and $D_{s}^{+}\rightarrow\pi^{+}\pi^{+}\pi^{-}$ based on the final state interaction is given in Sec. \ref{sec:Formalism}.
The numerical results are then presented in Sec. \ref{sec:Results}.
Finally, a brief summary is given in Sec. \ref{sec:Summary}.

\section{Formalism}\label{sec:Formalism}

This section concentrates on the theoretical formalism for the three-body decays $D_{s}^{+}\rightarrow \pi^{+}\pi^{+}\pi^{-}$ and $D^{+}\rightarrow \pi^{+}\pi^{+}\pi^{-}$.
First, the magnitude and phase of the $\pi^{+}\pi^{-}$ $S$-wave amplitude in the $D_{s}^{+}$ and $D^{+}$ decays are derived based on the final state interaction.
Second, the required scattering amplitudes of the five coupled channels $\pi^{+}\pi^{-}$, $\pi^{0}\pi^{0}$, $K^{+}K^{-}$, $K^{0}\bar{K}^{0}$, and $\eta\eta$ are calculated by the ChUA.

\subsection{The decay of $D_{s}^{+}\rightarrow \pi^{+}\pi^{+}\pi^{-}$}
\label{sec:A}

\begin{figure}[htbp]
\begin{minipage}{0.9\linewidth}
\centering
\includegraphics[width=1\linewidth,trim=140 550 190 120,clip]{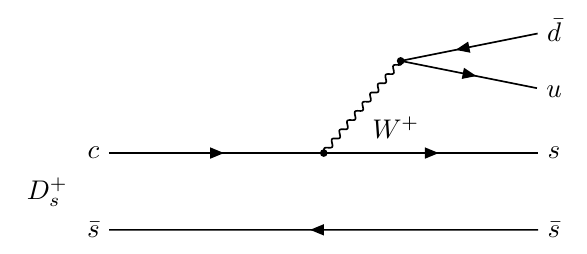} 
\label{fig:Feynman1}
\end{minipage}
\begin{minipage}{0.9\linewidth} 
\centering 
\includegraphics[width=1\linewidth,trim=140 550 190 120,clip]{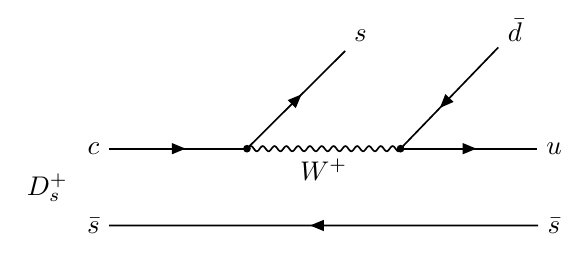} 
\label{fig:Feynman2}  
\end{minipage}
\caption{Diagrams for the $D_{s}^{+}\rightarrow\pi^{+}\pi^{+}\pi^{-}$ decay with external (above) and internal (below) $W$-emission mechanisms.}
\label{fig:Dsdecay}
\end{figure}

\begin{figure}[htbp]
\begin{minipage}{0.9\linewidth}
\centering
\includegraphics[width=1\linewidth,trim=150 580 170 120,clip]{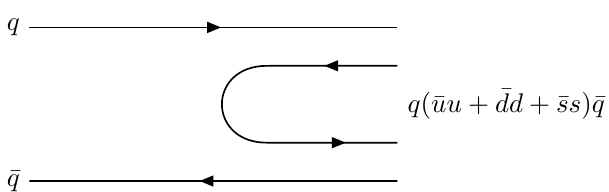} 
\end{minipage}
\caption{Procedure for the hadronization $q\bar{q}\rightarrow q(\bar{u}u+\bar{d}d+\bar{s}s)\bar{q}$.}
\label{fig:Hadronization}
\end{figure}

Although the final states in the $D_{s}^{+}\rightarrow \pi^{+}\pi^{+}\pi^{-}$ reaction are the same as that in the $D^{+}\rightarrow \pi^{+}\pi^{+}\pi^{-}$ reaction, the reaction mechanisms are different.
For the Cabibbo-favored process $D_{s}^{+}\rightarrow \pi^{+}\pi^{+}\pi^{-}$, the dominant mechanisms are the external and internal $W$-emission \cite{Chau:1982da,Chau:1987tk}, as depicted in Fig. \ref{fig:Dsdecay}.
For the external $W$-emission mechanism at the quark level, the $c$ quark of the initial $D_{s}^{+}$ meson can decay into a $W^{+}$ boson and a $s$ quark, while the $\bar{s}$ quark of $D_{s}^{+}$ remains a spectator.
Then the $W^{+}$ boson creates the $u$ and $\bar{d}$ quarks. 
Next, the $u\bar{d}$ ($s\bar{s}$) quark pair hadronises into $\pi^{+}$ ($\eta$) meson, with the other primary $s\bar{s}$ ($u\bar{d}$) pair simultaneously generating a meson pair, which is achieved by additional $\bar{q}q$ pairs generated from the vacuum, written as $\bar{u}u+\bar{d}d+\bar{s}s$, as shown in Fig. \ref{fig:Hadronization}.  
The formalism for the process can be written as follows

\begin{equation}
\begin{aligned}
|H_{D_{s}^{+}}^{(1)}\rangle
=&V_{P}V_{cs}V_{ud}\left[(u\bar{d}\rightarrow\pi^{+})|s(\bar{u}u+\bar{d}d+\bar{s}s)\bar{s}\rangle \right.\\ & \left. 
+(s\bar{s}\rightarrow\frac{-2}{\sqrt{6}}\eta)|u(\bar{u}u+\bar{d}d+\bar{s}s)\bar{d}\rangle\right] \\
=&V_{P}V_{cs}V_{ud}\left[\pi^{+}(M\cdot M)_{33}-\frac{2}{\sqrt{6}}\eta(M\cdot M)_{12}\right],
\end{aligned}
\label{eq:HDsa}
\end{equation}
where $V_{P}$ is the vertex factor of the weak interaction strength \cite{Liang:2015qva,Ahmed:2020qkv}, which is assumed as a real constant, and $V_{q_{1}q_{2}}$ is the element of the Cabibbo-Kobayashi-Maskawa (CKM) matrix for the transition of the quark $q_{1}\rightarrow q_{2}$.
In Eq. \eqref {eq:HDsa}, the factor $-2/\sqrt{6}$ of $\eta$ comes from the flavor component of $\eta$, i.e., $|\eta\rangle=|(u\bar{u}+d\bar{d}-2s\bar{s})/\sqrt{6}\rangle$.
The matrix $M$ for the $q\bar{q}$ elements is
\begin{equation}
\begin{aligned}
M=\left(\begin{array}{lll}{u \bar{u}} & {u \bar{d}} & {u \bar{s}} \\ {d \bar{u}} & {d \bar{d}} & {d \bar{s}} \\ {s \bar{u}} & {s \bar{d}} & {s \bar{s}}\end{array}\right).
\end{aligned}
\label{eq:M}
\end{equation}
Analogously, the formalism for the $D_{s}^{+}\rightarrow \pi^{+}\pi^{+}\pi^{-}$ decay of the internal $W$-emission mechanism can be written as
\begin{equation}
\begin{aligned}
|H_{D_{s}^{+}}^{(2)}\rangle
=&V_{P}\beta V_{cs}V_{ud}\left[(s\bar{d}\rightarrow\bar{K}^{0})|u(\bar{u}u+\bar{d}d+\bar{s}s)\bar{s}\rangle \right.\\ & \left.
+(u\bar{s}\rightarrow K^{+})|s(\bar{u}u+\bar{d}d+\bar{s}s)\bar{d}\rangle\right] \\
=&V_{P}\beta V_{cs}V_{ud}\left[\bar{K}^{0}(M\cdot M)_{13}+K^{+}(M\cdot M)_{32}\right],
\end{aligned}
\label{eq:HDsb}
\end{equation}
where $\beta$ is the value for the color suppression \cite{Duan:2020vye,Wei:2021usz}.

Furthermore, the matrix elements of $M$ can be written in terms of the physical mesons, which are
\begin{equation}
\begin{aligned}
P=\left(\begin{array}{ccc}{\frac{1}{\sqrt{2}} \pi^{0}+\frac{1}{\sqrt{6}} \eta} & {\pi^{+}} & {K^{+}} \\ {\pi^{-}} & {-\frac{1}{\sqrt{2}} \pi^{0}+\frac{1}{\sqrt{6}} \eta} & {K^{0}} \\ {K^{-}} & {\bar{K}^{0}} & {-\frac{2}{\sqrt{6}} \eta}\end{array}\right),
\end{aligned}
\label{eq:P}
\end{equation}
where we take $\eta \equiv \eta_{8}$ as in Ref. \cite{Liang:2014tia}. 
With the corresponding relations between the matrices $M$ and $P$, 
the hadronization processes at the quark level in Eqs. (\ref{eq:HDsa}) and (\ref{eq:HDsb}) can be rewritten at the hadron level in terms of two pseudoscalar mesons. 
In this way, we can determine the relevant components of the two meson pairs as follows
\begin{equation}
\begin{aligned}
|H_{D_{s}^{+}}^{(1)}\rangle
=&V_{P}V_{cs}V_{ud}\left[\pi^{+}(K^{+}K^{-}+K^{0}\bar{K}^{0}+\frac{2}{3}\eta\eta) \right.\\ & \left.
-\frac{2}{\sqrt{6}}\eta(\frac{2}{\sqrt{6}}\eta\pi^{+})\right],
\end{aligned}
\label{eq:HDs1}
\end{equation}
\begin{equation}
\begin{aligned}
|H_{D_{s}^{+}}^{(2)}\rangle
=&V_{P}\beta V_{cs}V_{ud}\left[\bar{K}^{0}(\pi^{+}K^{0}) 
+K^{+}(\pi^{+}K^{-}) \right].
\end{aligned}
\label{eq:HDs2}
\end{equation}
\begin{figure*}[htbp]
\begin{minipage}{0.44\linewidth}
\centering
\includegraphics[width=1\linewidth,trim=150 540 180 130,clip]{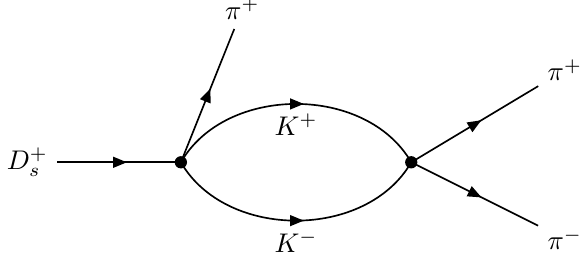} 
\label{fig:Scatter1}
\end{minipage}
\quad
\quad
\begin{minipage}{0.44\linewidth}
\centering
\includegraphics[width=1\linewidth,trim=150 540 180 130,clip]{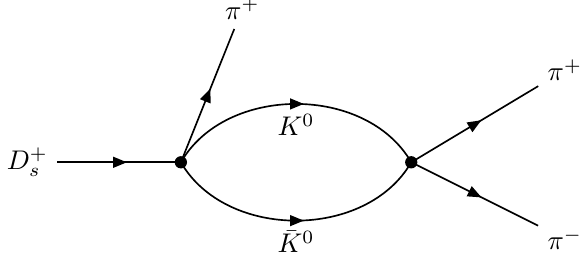} 
\label{fig:Scatter2}
\end{minipage}
\caption{Mechanisms of the $S$-wave final state interactions in the $D_{s}^{+}$ decay.}
\label{fig:DsScatter}
\end{figure*}
Note that, we only keep the terms that contribute to the final states $\pi^{+}\pi^{+}\pi^{-}$ in Eqs. \eqref{eq:HDs1} and \eqref{eq:HDs2}.
Then, we can obtain the total contributions in the
$S$-wave from the Feynman diagrams in Fig. \ref{fig:Dsdecay},
\begin{equation}
\begin{aligned}
|H_{D_{s}^{+}}\rangle
=&|H_{D_{s}^{+}}^{(1)}\rangle+|H_{D_{s}^{+}}^{(2)}\rangle \\
=&V_{P}(1+\beta) V_{cs}V_{ud}\left(\pi^{+}K^{+}K^{-}+\pi^{+}K^{0}\bar{K}^{0}\right).
\end{aligned}
\label{eq:HDs}
\end{equation}
From Eq. \eqref{eq:HDs}, it can be seen that there are only $K^{+}K^{-}$ and $K^{0}\bar{K}^{0}$ channels are produced in the $D_{s}^{+}$ decay.
There is no contribution from the tree level, all the final states $\pi^{+}\pi^{+}\pi^{-}$ are obtained by two-body rescattering.
As we know from the ChUA, the $f_{0}(500)$ state is contributed by the $\pi\pi$ channel, while the $f_{0}(980)$ state is mainly contributed by the $K\bar{K}$ channel, discussed in more detailedly in Ref. \cite{Ahmed:2020kmp}.
Therefore, we can expect the $f_{0}(980)$ to be produced in the $D_{s}^{+}$ decay, whereas $f_{0}(500)$ is not, as the experimental suggestions \cite{BaBar:2008nlp,BESIII:2021jnf,LHCb:2022pjv}.
Note that $(K^{+}K^{-}+K^{0}\bar{K}^{0})$ has only components of isospin $I=0$, but no components of $I=1$, so there is no contribution from the $a_{0}(980)$ resonance here.
Once the final states are hadronized after the weak decay process, they can undergo further interaction, as shown in Fig. \ref{fig:DsScatter}, where the final state $\pi^{+}\pi^{-}$ can be obtained through the final state interactions.
The amplitudes for these final state productions and their interactions can be written as
\begin{equation}
\begin{aligned}
t_{D_{s}^{+}\rightarrow\pi^{+}\pi^{+}\pi^{-}}
=\mathcal{A}_{S-wave}(s_{13})+(s_{13}\leftrightarrow s_{23}),
\end{aligned}
\label{eq:Dsamplitudes}
\end{equation}
which the $s_{ij}$ is the square of the energy of two particles in the center-of-mass frame, the lower indices $i,j=1,2,3$ denote the three final states $\pi^{+}$(1), $\pi^{+}$(2), and $\pi^{-}$(3), respectively.
The symbol $s_{13}\leftrightarrow s_{23}$ means exchanging the $s_{13}$ and $s_{23}$ in the amplitude, which represents the symmetry of identical particles $\pi^{+}\pi^{+}$ in the final states.
And we also have
\begin{equation}
\begin{aligned}
\mathcal{A}_{S-wave}(s_{13})
=&\mathcal{C}
\left[G_{K^{+}K^{-}}(s_{13})
T_{K^{+}K^{-} \rightarrow \pi^{+}\pi^{-}}(s_{13})
\right.\\ & \left.+G_{K^{0}\bar{K}^{0}}(s_{13})
T_{K^{0}\bar{K}^{0} \rightarrow \pi^{+}\pi^{-}}(s_{13})\right],
\end{aligned}
\label{eq:DsA}
\end{equation}
where $C$ is a global factor that includes the production vertex $V_{P}$, $1+\beta$, and the elements of the CKM matrix, as well as the normalization factor used to match the experimental data. 
Besides, $G$ and $T$ are the loop functions and the scattering amplitudes, respectively. 
These terms will be introduced later in subsection \ref{sec:C}.

\subsection{The decay of $D^{+}\rightarrow \pi^{+}\pi^{+}\pi^{-}$}
\label{sec:B}

\begin{figure*}[htbp]
\begin{minipage}{0.44\linewidth}
\centering
\includegraphics[width=1\linewidth,trim=140 550 190 120,clip]{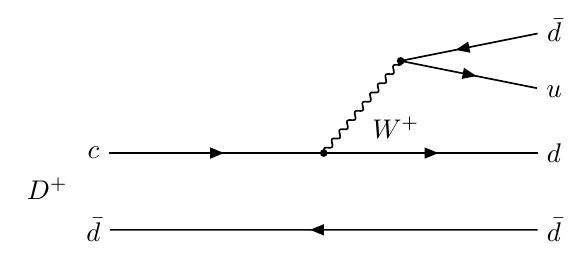} 
\label{fig:Feynman3}
\end{minipage}
\begin{minipage}{0.44\linewidth} 
\centering 
\includegraphics[width=1\linewidth,trim=140 550 190 120,clip]{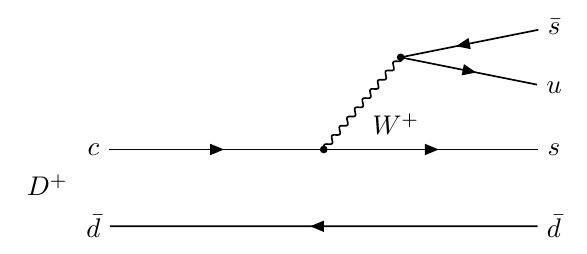} 
\label{fig:Feynman4}  
\end{minipage}
\begin{minipage}{0.44\linewidth} 
\centering 
\includegraphics[width=1\linewidth,trim=140 550 190 120,clip]{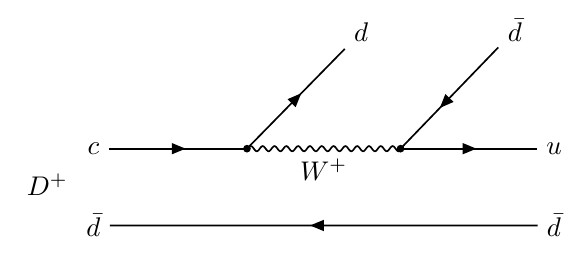} 
\label{fig:Feynman5}  
\end{minipage}
\begin{minipage}{0.44\linewidth} 
\centering 
\includegraphics[width=1\linewidth,trim=140 550 190 120,clip]{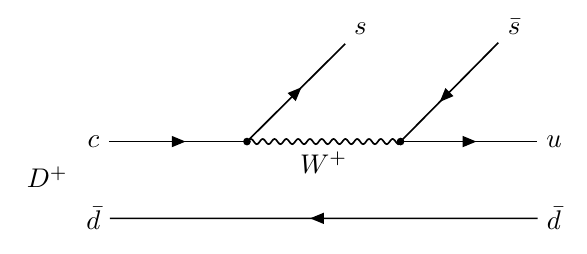} 
\label{fig:Feynman6}  
\end{minipage}
\caption{Diagrams for the $D^{+}\rightarrow\pi^{+}\pi^{+}\pi^{-}$ decay with external (above row) and internal (below row) $W$-emission mechanisms.}
\label{fig:Ddecay}
\end{figure*}

\begin{figure*}[htbp]
\begin{minipage}{0.44\linewidth}
\centering
\includegraphics[width=1\linewidth,trim=140 550 210 130,clip]{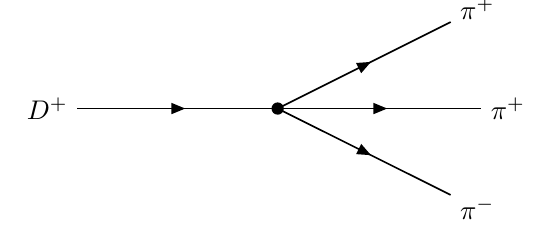} 
\label{fig:Scatter3}
\end{minipage}
\quad
\quad
\begin{minipage}{0.44\linewidth}
\centering
\includegraphics[width=1\linewidth,trim=150 540 180 130,clip]{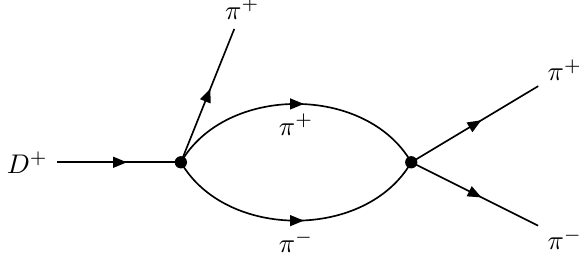} 
\label{fig:Scatter4}
\end{minipage}
\begin{minipage}{0.44\linewidth}
\centering
\includegraphics[width=1\linewidth,trim=150 540 180 130,clip]{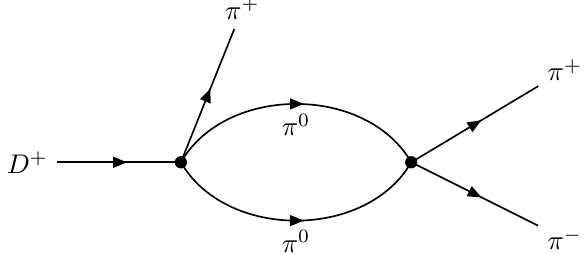} 
\label{fig:Scatter5}
\end{minipage}
\quad
\quad
\begin{minipage}{0.44\linewidth}
\centering
\includegraphics[width=1\linewidth,trim=150 540 180 130,clip]{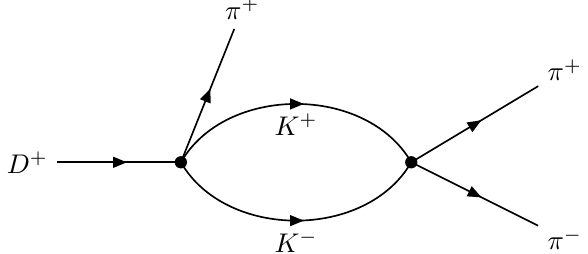} 
\label{fig:Scatter6}
\end{minipage}
\begin{minipage}{0.44\linewidth}
\centering
\includegraphics[width=1\linewidth,trim=150 540 180 130,clip]{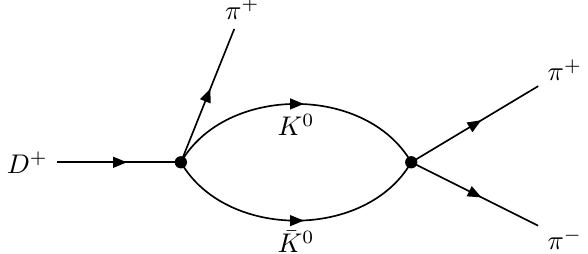} 
\label{fig:Scatter7}
\end{minipage}
\quad
\quad
\begin{minipage}{0.44\linewidth}
\centering
\includegraphics[width=1\linewidth,trim=150 540 180 130,clip]{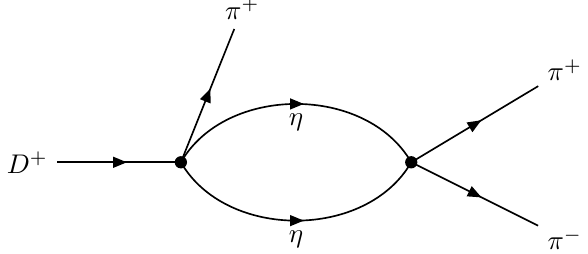} 
\label{fig:Scatter8}
\end{minipage}
\caption{Mechanisms of the $S$-wave final state interactions in the $D^{+}$ decay.}
\label{fig:DScatter}
\end{figure*}

In this subsection, we present the theoretical formalism for the $D^{+}\rightarrow \pi^{+}\pi^{+}\pi^{-}$ decay.
The dominant external and internal $W$-emission mechanisms are taken into account, and the corresponding Feynman diagrams are shown in Fig. \ref{fig:Ddecay}.
Following the $D_{s}^{+}$ decay discussed above, we can derive the formula for the weak decay of $D^{+}$ as follows
\begin{equation}
\begin{aligned}
|H_{D^{+}}^{(1)}\rangle
=&V_{P}V_{cd}V_{ud}\left[(u\bar{d}\rightarrow\pi^{+})|d(\bar{u}u+\bar{d}d+\bar{s}s)\bar{d}\rangle \right.\\ & \left.
+(d\bar{d}\rightarrow\frac{-1}{\sqrt{2}}\pi^{0})|u(\bar{u}u+\bar{d}d+\bar{s}s)\bar{d}\rangle \right.\\ & \left.
+(d\bar{d}\rightarrow\frac{1}{\sqrt{6}}\eta)|u(\bar{u}u+\bar{d}d+\bar{s}s)\bar{d}\rangle \right]\\
=&V_{P}V_{cd}V_{ud}\left[\pi^{+}(M\cdot M)_{22}-\frac{1}{\sqrt{2}}\pi^{0}(M\cdot M)_{12}
\right.\\ & \left.+\frac{1}{\sqrt{6}}\eta(M\cdot M)_{12}\right],
\end{aligned}
\label{eq:HDa}
\end{equation}
\begin{equation}
\begin{aligned}
|H_{D^{+}}^{(2)}\rangle
=&V_{P}^{'}V_{cs}V_{us}\left[(u\bar{s}\rightarrow K^{+})|s(\bar{u}u+\bar{d}d+\bar{s}s)\bar{d}\rangle \right.\\ & \left. 
+(s\bar{d}\rightarrow \bar{K}^{0})|u(\bar{u}u+\bar{d}d+\bar{s}s)\bar{s}\rangle \right]\\
=&V_{P}^{'}V_{cs}V_{us}\left[K^{+}(M\cdot M)_{32}+\bar{K}^{0}(M\cdot M)_{13}\right],
\end{aligned}
\label{eq:HDb}
\end{equation}
\begin{equation}
\begin{aligned}
|H_{D^{+}}^{(3)}\rangle
=&V_{P}\beta V_{cd}V_{ud}\left[(d\bar{d}\rightarrow\frac{-1}{\sqrt{2}}\pi^{0})|u(\bar{u}u+\bar{d}d+\bar{s}s)\bar{d}\rangle \right.\\ & \left.
+(d\bar{d}\rightarrow\frac{1}{\sqrt{6}}\eta)|u(\bar{u}u+\bar{d}d+\bar{s}s)\bar{d}\rangle \right.\\ & \left.
+(u\bar{d}\rightarrow\pi^{+})|d(\bar{u}u+\bar{d}d+\bar{s}s)\bar{d}\rangle \right]\\ 
=&V_{P}\beta V_{cd}V_{ud}\left[-\frac{1}{\sqrt{2}}\pi^{0}(M\cdot M)_{12}+\frac{1}{\sqrt{6}}\eta(M\cdot M)_{12} \right.\\ & \left.
+\pi^{+}(M\cdot M)_{22}\right],
\end{aligned}
\label{eq:HDc}
\end{equation}
\begin{equation}
\begin{aligned}
|H_{D^{+}}^{(4)}\rangle
=&V_{P}^{'}\beta V_{cs}V_{us}\left[(s\bar{s}\rightarrow \frac{-2}{\sqrt{6}}\eta)|u(\bar{u}u+\bar{d}d+\bar{s}s)\bar{d}\rangle \right.\\ & \left. 
+(u\bar{d}\rightarrow \pi^{+})|s(\bar{u}u+\bar{d}d+\bar{s}s)\bar{s}\rangle \right]\\
=&V_{P}^{'}\beta V_{cs}V_{us}\left[-\frac{2}{\sqrt{6}}\eta(M\cdot M)_{12}+\pi^{+}(M\cdot M)_{33}\right],
\end{aligned}
\label{eq:HDd}
\end{equation}
where the factor $-1/\sqrt{2}$ of $\pi^{0}$ comes from the flavor component of $\pi^{0}$, i.e., $|\pi^{0}\rangle=|(u\bar{u}-d\bar{d})/\sqrt{2}\rangle$.
The Eqs. \eqref{eq:HDa} and \eqref{eq:HDb} are contributions from the external $W$-emission, while Eqs. \eqref{eq:HDc} and \eqref{eq:HDd} are contributions from the internal $W$-emission.
The hadronization processes at the quark level in Eqs. \eqref{eq:HDa}–\eqref{eq:HDd} can be transferred to the hadron level in terms of two pseudoscalar mesons, given by
\begin{equation}
\begin{aligned}
|H_{D^{+}}^{(1)}\rangle
=&V_{P}V_{cd}V_{ud}\left[\pi^{+}(\pi^{+}\pi^{-}+\frac{1}{2}\pi^{0}\pi^{0}+K^{0}\bar{K}^{0}+\frac{1}{6}\eta\eta) \right.\\ & \left.
+\frac{1}{\sqrt{6}}\eta(\frac{2}{\sqrt{6}}\eta\pi^{+})\right],
\end{aligned}
\label{eq:HD1}
\end{equation}
\begin{equation}
\begin{aligned}
|H_{D^{+}}^{(2)}\rangle
=&V_{P}^{'}V_{cs}V_{us}\left[K^{+}(K^{-}\pi^{+})+\bar{K}^{0}(K^{0}\pi^{+})\right],
\end{aligned}
\label{eq:HD2}
\end{equation}
\begin{equation}
\begin{aligned}
|H_{D^{+}}^{(3)}\rangle
=&V_{P}\beta V_{cd}V_{ud}\left[\frac{1}{\sqrt{6}}\eta(\frac{2}{\sqrt{6}}\eta\pi^{+})
+\pi^{+}(\pi^{+}\pi^{-}+\frac{1}{2}\pi^{0}\pi^{0}\right.\\ & \left.+K^{0}\bar{K}^{0}+\frac{1}{6}\eta\eta)\right],
\end{aligned}
\label{eq:HD3}
\end{equation}
\begin{equation}
\begin{aligned}
|H_{D^{+}}^{(4)}\rangle
=&V_{P}^{'}\beta V_{cs}V_{us}\left[-\frac{2}{\sqrt{6}}\eta(\frac{2}{\sqrt{6}}\eta\pi^{+})+\pi^{+}(K^{+}K^{-}+K^{0}\bar{K}^{0}\right.\\ & \left.+\frac{2}{3}\eta\eta)\right].
\end{aligned}
\label{eq:HD4}
\end{equation}

The total contributions in the $S$-wave are the sum of these four formalisms mentioned above
\begin{equation}
\begin{aligned}
|H_{D^{+}}\rangle
=&|H_{D^{+}}^{(1)}\rangle+|H_{D^{+}}^{(2)}\rangle+|H_{D^{+}}^{(3)}\rangle+|H_{D^{+}}^{(4)}\rangle \\
=&(1+\beta) \left[V_{P}V_{cd}V_{ud}(\pi^{+}\pi^{+}\pi^{-}+\frac{1}{2}\pi^{+}\pi^{0}\pi^{0}+\pi^{+}K^{0}\bar{K}^{0}  \right.\\ & \left.
+\frac{1}{2}\pi^{+}\eta\eta)+V_{P}^{'}V_{cs}V_{us}(\pi^{+}K^{+}K^{-}+\pi^{+}K^{0}\bar{K}^{0}) \right].
\end{aligned}
\label{eq:HD}
\end{equation}
In the $D^{+}\rightarrow \pi^{+}\pi^{+}\pi^{-}$ process, there are five coupled channels that contributing to the decay, so it is implied that both $f_{0}(500)$ and $f_{0}(980)$ states exist.
Unlike the $D_{s}^{+}$ decay, the contributions from the tree level for $D^{+}$ are not forbidden.
In Eq. \eqref{eq:HD}, the terms $\pi^{+}\pi^{-}$, $\pi^{0}\pi^{0}$, $\eta\eta$, and ($K^{+}K^{-}+K^{0}\bar{K}^{0}$) have only components of isospin $I=0$, while the extra $K^{0}\bar{K}^{0}$ has components of both $I=0$ and $I=1$.
So it can be expected that the contribution of $f_{0}(980)$ will be much larger than that of $a_{0}(980)$, and we will not emphasize the $a_{0}(980)$ state in the numerical results section later.
After the two-body interaction rescattering procedure, as shown in Fig. \ref{fig:DScatter}, the total amplitudes in the $S$-wave are
\begin{equation}
\begin{aligned}
t_{D^{+}\rightarrow\pi^{+}\pi^{+}\pi^{-}}
=\mathcal{A}_{S-wave}(s_{13})+(s_{13}\leftrightarrow s_{23})
\end{aligned}
\label{eq:Damplitudes}
\end{equation}
with
\begin{equation}
\begin{aligned}
\mathcal{A}_{S-wave}(s_{13})
=&\mathcal{D}_{1}\left[1+G_{\pi^{+}\pi^{-}}(s_{13})T_{\pi^{+}\pi^{-} \rightarrow \pi^{+}\pi^{-}}(s_{13})\right.\\ & \left.
+\frac{1}{2}G_{\pi^{0}\pi^{0}}(s_{13})T_{\pi^{0}\pi^{0} \rightarrow \pi^{+}\pi^{-}}(s_{13})\right.\\ & \left.
+G_{K^{0}\bar{K}^{0}}(s_{13})T_{K^{0}\bar{K}^{0} \rightarrow \pi^{+}\pi^{-}}(s_{13})\right.\\ & \left.
+\frac{1}{2}G_{\eta\eta}(s_{13})
T_{\eta\eta \rightarrow \pi^{+}\pi^{-}}(s_{13})\right] \\
&+\mathcal{D}_{2}\left[G_{K^{+}K^{-}}(s_{13}) T_{K^{+}K^{-} \rightarrow \pi^{+}\pi^{-}}(s_{13}) \right.\\ & \left.
+G_{K^{0}\bar{K}^{0}}(s_{13}) T_{K^{0}\bar{K}^{0} \rightarrow \pi^{+}\pi^{-}}(s_{13})\right],
\end{aligned}
\label{eq:DA}
\end{equation}
where the factor $\mathcal{D}_{1}$ ($\mathcal{D}_{2}$) includes $V_{P}$ ($V_{P}^{'}$), $1+\beta$, the elements of the CKM matrix, and the normalization factor for the experimental data.
Note that there is a factor of $2$ in the terms related to the identical particles $\pi^{0}\pi^{0}$ and $\eta\eta$ in Eq. \eqref{eq:DA}, which has been cancelled by the $1/2$ factor in their propagators, for further discussion see Ref. \cite{Liang:2015qva}.

\subsection{The chiral unitary approach for the $f_{0}(500)$ and $f_{0}(980)$}
\label{sec:C}

In the chiral unitary approach, the states of $f_{0}(500)$ and $f_{0}(980)$ are dynamically generated in the coupled channel interactions.
For the isospin $I=0$, there are five coupled channels $\pi^{+}\pi^{-}$ (1), $\pi^{0}\pi^{0}$ (2), $K^{+}K^{-}$ (3), $K^{0}\bar{K}^{0}$ (4), and $\eta\eta$ (5) in the physical basis.
Using the lowest order chiral Lagrangian, the interaction potentials in the $S$-wave are given by \cite{Liang:2014tia,Ahmed:2020qkv}
\begin{equation}
\begin{aligned}
&V_{11}=-\frac{1}{2 f^{2}} s, \quad V_{12}=-\frac{1}{\sqrt{2} f^{2}}\left(s-m_{\pi}^{2}\right), \quad V_{13}=-\frac{1}{4 f^{2}} s ,\\
&V_{14}=-\frac{1}{4 f^{2}} s, \quad V_{15}=-\frac{1}{3 \sqrt{2} f^{2}} m_{\pi}^{2}, \quad V_{22}=-\frac{1}{2 f^{2}} m_{\pi}^{2} ,\\
&V_{23}=-\frac{1}{4 \sqrt{2} f^{2}} s, \quad V_{24}=-\frac{1}{4 \sqrt{2} f^{2}} s, \quad V_{25}=-\frac{1}{6 f^{2}} m_{\pi}^{2} ,\\
&V_{33}=-\frac{1}{2 f^{2}} s, \quad V_{34}=-\frac{1}{4 f^{2}} s ,\\
&V_{35}=-\frac{1}{12 \sqrt{2} f^{2}}\left(9 s-6 m_{\eta}^{2}-2 m_{\pi}^{2}\right), \quad V_{44}=-\frac{1}{2 f^{2}} s ,\\
&V_{45}=-\frac{1}{12 \sqrt{2} f^{2}}\left(9 s-6 m_{\eta}^{2}-2 m_{\pi}^{2}\right) ,\\
&V_{55}=-\frac{1}{18 f^{2}}\left(16 m_{K}^{2}-7 m_{\pi}^{2}\right),
\end{aligned}
\end{equation}
where $f$ is the pion decay constant, and we take $f=0.093$ GeV.

Then one can solve the Bethe-Salpeter equation factorized on shell, to obtain the two-body scattering amplitudes of the coupled channels \cite{Oller:1997ti,Oset:1997it,Oller:1997ng}
\begin{equation}
\begin{aligned}
T = [1-VG]^{-1}V.
\end{aligned}
\label{eq:BSE}
\end{equation}
Furthermore, $G$ is a diagonal matrix consisting of the loop function of two intermediate meson propagators, which can be written as
\begin{equation}
\begin{aligned} 
G_{ii}=i \int \frac{d^{4} q}{(2 \pi)^{4}}\frac{1}{(P-q)^{2}-m_{1}^{2}+i \epsilon} \frac{1}{q^{2}-m_{2}^{2}+i \epsilon},
\end{aligned}
\label{eq:G}
\end{equation}
where $P=p_{1}+p_{2}$ is the total momentum of the two mesons coupled system, and $m_{1}$, $m_{2}$ are the masses of the two intermediate mesons in the loop.
Note that this magnitude is logarithmically divergent and there are two methods for solving this singular integral.
The first one is the three-momentum cutoff method \cite{Oller:1997ti},
\begin{equation}
\begin{aligned}
G _ { ii } ( s ) = \int _ { 0 } ^ { q _ { \max } } \frac { q ^ { 2 } d q } { ( 2 \pi ) ^ { 2 } } \frac { \omega _ { 1 } + \omega _ { 2 } } { \omega _ { 1 } \omega _ { 2 } \left[ s - \left( \omega _ { 1 } + \omega _ { 2 } \right) ^ { 2 } + i \epsilon \right] },
\end{aligned}
\label{eq:GCO}
\end{equation}
where $\omega_{i}(\vec{q})=\sqrt{\vec{q}\,^2+m_{i}^2}$, and $q_{max}$ is the only free parameter. 
The other one is the dimensional regularization method \cite{Oller:1998zr,Oller:2000fj,Gamermann:2006nm,Alvarez-Ruso:2010rqm,Guo:2016zep}, given by 
\begin{equation}
\begin{aligned}
G_{ii}(s)=& \frac{1}{16 \pi^{2}}\left\{a_{i}(\mu)+\ln \frac{m_{1}^{2}}{\mu^{2}}+\frac{m_{2}^{2}-m_{1}^{2}+s}{2 s} \ln \frac{m_{2}^{2}}{m_{1}^{2}}\right.\\
&+\frac{q_{cmi}(s)}{\sqrt{s}}\left[\ln \left(s-\left(m_{2}^{2}-m_{1}^{2}\right)+2 q_{cmi}(s) \sqrt{s}\right)\right.\\
&+\ln \left(s+\left(m_{2}^{2}-m_{1}^{2}\right)+2 q_{cmi}(s) \sqrt{s}\right) \\
&-\ln \left(-s-\left(m_{2}^{2}-m_{1}^{2}\right)+2 q_{cmi}(s) \sqrt{s}\right) \\
&\left.\left.-\ln \left(-s+\left(m_{2}^{2}-m_{1}^{2}\right)+2 q_{cmi}(s) \sqrt{s}\right)\right]\right\},
\end{aligned}
\label{eq:GDR}
\end{equation}
which also has only one free parameter, the regularization scale $\mu$, and the subtraction constant $a_{i}(\mu)$ depends on the chosen $\mu$, the discussion of their relationship can be found in Refs. \cite{Oller:2000fj,Guo:2018tjx}.
The $q_{cmi}(s)$ is three-momentum of the particle in the center-of-mass frame
\begin{equation}
\begin{aligned}
q_{cmi}(s)=\frac{\lambda^{1 / 2}\left(s, m_{1}^{2}, m_{2}^{2}\right)}{2 \sqrt{s}},
\end{aligned}
\label{eq:qcmi}
\end{equation}
with the K\"all\'en triangle function $\lambda(a, b, c)=a^{2}+b^{2}+c^{2}-2(a b+a c+b c)$.
In the present work. we use the dimensional regularization method.
In fact, whichever method we choose will not affect the conclusions.
The values of the regularization scale $\mu$ and the subtraction constant $a_{i}(\mu)$ in loop functions will be discussed in detail in Sec. \ref{sec:Results}.

\section{Numerical results}\label{sec:Results}

In our formalism, the first thing to do is to determine the subtraction constants $a_{i}(\mu)$ for the different coupled channels in the loop functions.
As mentioned in the previous section, it depends on the value of the regularization scale $\mu$. 
We combine two different renormalization schemes to make them equal at the threshold for a given channel \cite{Oset:2001cn,Montana:2022inz}, and thus we obtain
\begin{equation}
\begin{aligned}
a_{i}(\mu)=16\pi^{2}[G^{CO}(s_{thr},q_{max})-G^{DR}(s_{thr},\mu)],
\end{aligned}
\label{eq:ai}
\end{equation}
where we take $q_{max}=\mu$ since they have the same unit, $G^{CO}$ and $G^{DR}$ are given by Eqs. \eqref{eq:GCO} and \eqref{eq:GDR}, respectively.
This helps us to ensure that the results of these two methods are similar.
For the regularization scale $\mu$, we adopt two different schemes: In scheme I, we take $\mu=0.6$ GeV which is done in Refs. \cite{Liang:2014tia,Dias:2016gou,Duan:2020vye}, while in scheme II, we treat $\mu$ as a free parameter and determine it by fitting experimental data.
The masses of pseudoscalar and charmed mesons in this paper are taken from the Particle Data Group \cite{ParticleDataGroup:2024cfk}.

\begin{table}[htbp]
\centering
\renewcommand\tabcolsep{2.0mm}
\renewcommand{\arraystretch}{1.50}
\caption{Values of the parameters from the fit in the $D_{s}^{+}$ decay.}
\begin{tabular*}{86mm}{@{\extracolsep{\fill}}l|cc}
\toprule[1.00pt]
\toprule[1.00pt]
Parameter&$\mu$&$\mathcal{C}$\\
\hline
Fit I&$0.6$ GeV (fixed)&$23.89\pm0.14$\\
\hline
Fit II &$0.561\pm0.001$ GeV&$28.75\pm0.22$\\
\bottomrule[1.00pt]
\bottomrule[1.00pt]
\end{tabular*}
\label{tab:DsParameter1}
\end{table}

\begin{figure}[htbp]
\begin{minipage}{0.8\linewidth}
\centering
\includegraphics[width=1\linewidth,trim=0 0 0 0,clip]{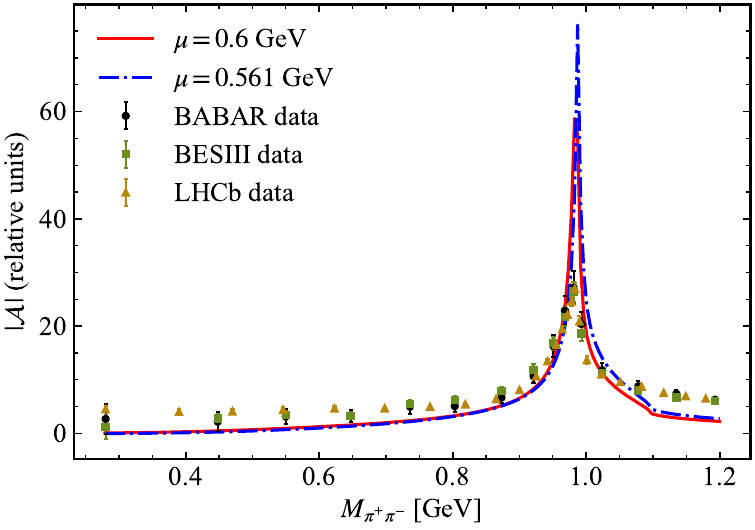} 
\label{fig:ADs}
\end{minipage}
\begin{minipage}{0.8\linewidth}
\centering
\includegraphics[width=1\linewidth,trim=0 0 0 0,clip]{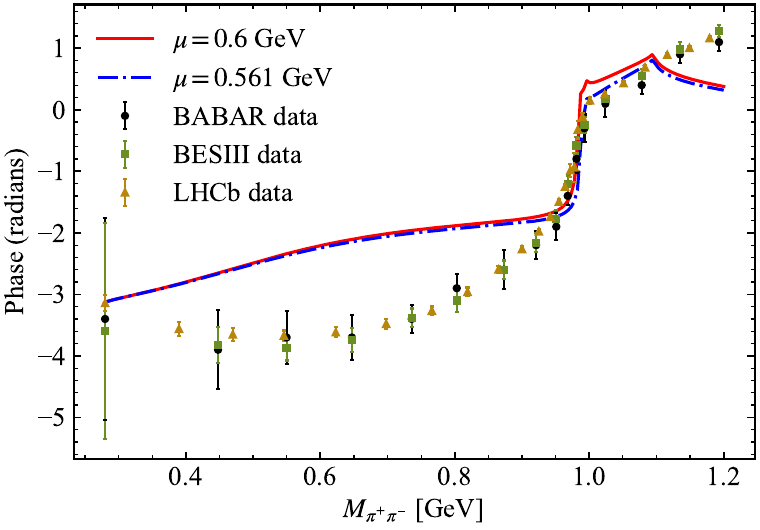} 
\label{fig:PDs}
\end{minipage}
\caption{The magnitude (above) and phase (below) of the $\pi^{+}\pi^{-}$ $S$-wave amplitude in the $D_{s}^{+}\rightarrow \pi^{+}\pi^{+}\pi^{-}$ decay. The points are the experimental data measured by the BABAR \cite{BaBar:2008nlp}, BESIII \cite{BESIII:2021jnf}, and LHCb \cite{LHCb:2022pjv} Collaborations, respectively.}
\label{fig:DsAandP}
\end{figure}

Firstly, we fit the experimental data of the $D_{s}^{+}\rightarrow \pi^{+}\pi^{+}\pi^{-}$ decay, and the parameters are listed in Table \ref{tab:DsParameter1}.
In Fit I, there is only one parameter, the global factor $\mathcal{C}$, since it does not contribute to the phase, we only fit the magnitude of the amplitude.
For the Fit II, we add an additional parameter $\mu$ which is related to the phase, so the fit combine magnitude and phase together.
The best results are obtained by a combined fit with the experimental data from the BABAR \cite{BaBar:2008nlp}, BESIII \cite{BESIII:2021jnf}, and LHCb \cite{LHCb:2022pjv} Collaborations which are shown in Fig. \ref{fig:DsAandP}.
Note that the ChUA is meaningful for regions with $\sqrt{s}$ less than $1.2$ GeV in the $K\bar{K}$ coupled channel interactions, as discussed in Refs. \cite{Oller:1997ti,Debastiani:2016ayp}. 
Therefore, the data below $1.2$ GeV were retained, and we are only interested in the energy regions of the $f_{0}(500)$ and $f_{0}(980)$ states.

In Fig. \ref{fig:DsAandP}, a clear peak structure appears around $0.98$ GeV in the magnitude of $\pi^{+}\pi^{-}$ amplitude, which is the $f_{0}(980)$ state dynamically generated in the ChUA.
There is no signal from the $f_{0}(500)$ state, which is consistent with the analysis of Eq. \eqref{eq:HDs} in Sec. \ref{sec:A}.
Note that our results are relatively small compared to experimental data in the regions below $0.8$ GeV and above $1.1$ GeV, while relatively large around $1.0$ GeV, which will be discussed later.
For the phase of the $\pi^{+}\pi^{-}$ $S$-wave amplitude, our results describe the experimental data well at around $1.0$ GeV, but are poorly in the $0.4-0.8$ GeV range.
It should be pointed out that there are no free parameters in Fit I for the calculation of the phase in the $D_{s}^{+}\rightarrow\pi^{+}\pi^{+}\pi^{-}$ decay.
Furthermore, there is little difference between the two schemes. 
This is also the case for the $D^{+}$ decay process.

\begin{table}[htbp]
\centering
\renewcommand\tabcolsep{2.0mm}
\renewcommand{\arraystretch}{1.50}
\caption{Values of the parameters from the fit in the $D^{+}$ decay.}
\begin{tabular*}{86mm}{@{\extracolsep{\fill}}l|ccc}
\toprule[1.00pt]
\toprule[1.00pt]
Parameter&$\mu$&$\mathcal{D}_{1}$&$\mathcal{D}_{2}$\\
\hline
Fit I&$0.6$ GeV (fixed)&$12.22\pm0.14$&$-12.05\pm0.15$\\
\hline
Fit II &$0.525\pm0.006$ GeV&$12.05\pm0.15$&$-16.74\pm0.43$\\
\bottomrule[1.00pt]
\bottomrule[1.00pt]
\end{tabular*}
\label{tab:DParameter1}
\end{table}

\begin{figure}[htbp]
\begin{minipage}{0.8\linewidth}
\centering
\includegraphics[width=1\linewidth,trim=0 0 0 0,clip]{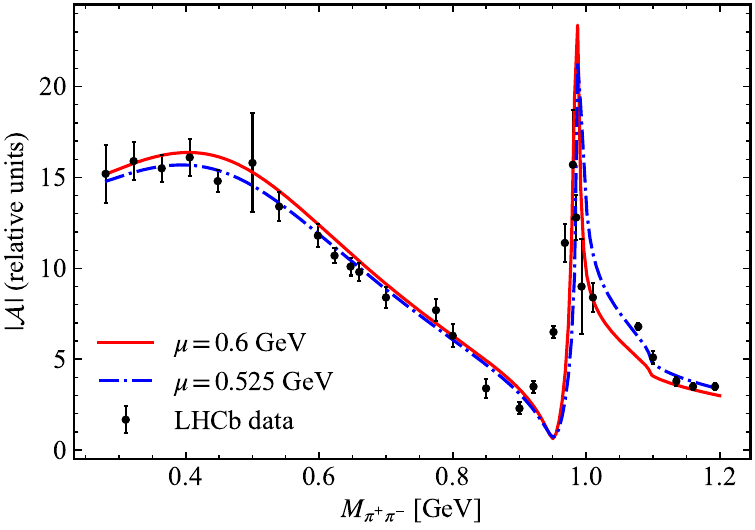} 
\label{fig:ADp}
\end{minipage}
\begin{minipage}{0.8\linewidth}
\centering
\includegraphics[width=1\linewidth,trim=0 0 0 0,clip]{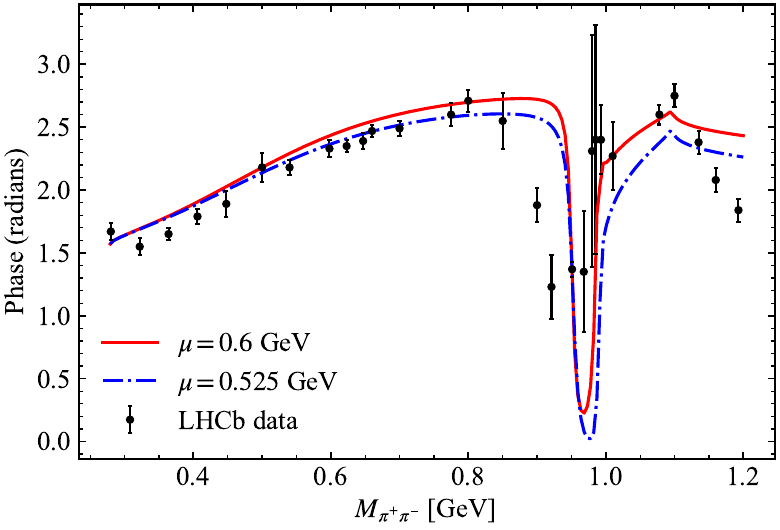} 
\label{fig:PDp}
\end{minipage}
\caption{The magnitude (above) and phase (below) of the $\pi^{+}\pi^{-}$ $S$-wave amplitude in the $D^{+}\rightarrow \pi^{+}\pi^{+}\pi^{-}$ decay. The points are the experimental data measured by the LHCb Collaboration \cite{LHCb:2022lja}.}
\label{fig:DAandP}
\end{figure}

\begin{figure*}[htbp]
\begin{minipage}{0.33\linewidth}
\centering
\includegraphics[width=1\linewidth,trim=0 0 0 0,clip]{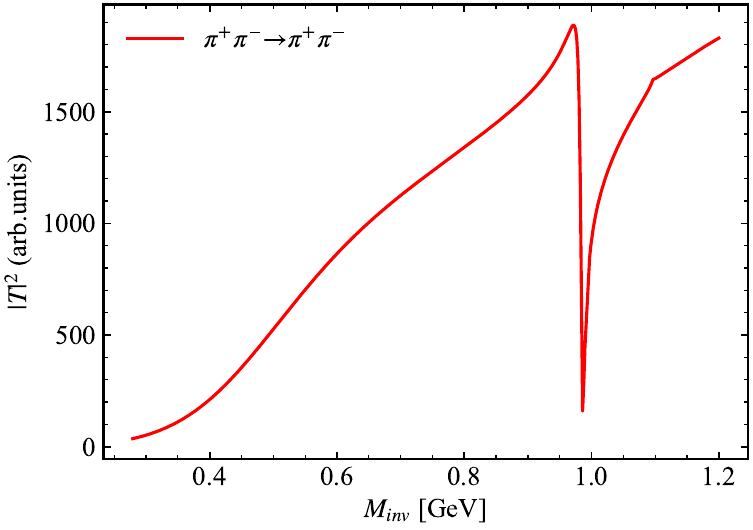} 
\label{fig:T11}
\end{minipage}
\begin{minipage}{0.33\linewidth}
\centering
\includegraphics[width=1\linewidth,trim=0 0 0 0,clip]{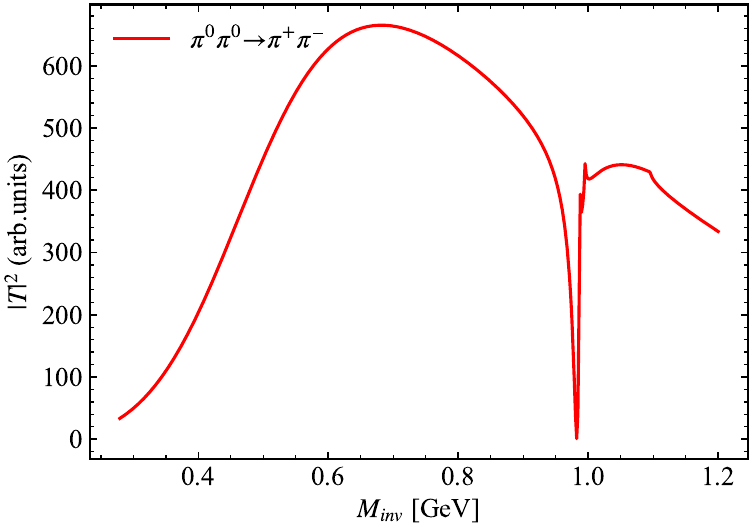} 
\label{fig:T21}
\end{minipage}
\begin{minipage}{0.33\linewidth}
\centering
\includegraphics[width=1\linewidth,trim=0 0 0 0,clip]{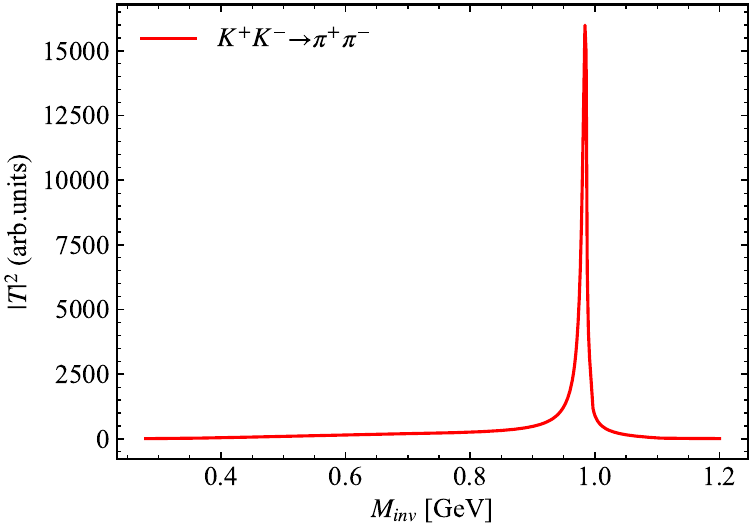} 
\label{fig:T31}
\end{minipage}
\quad
\quad
\begin{minipage}{0.33\linewidth}
\centering
\includegraphics[width=1\linewidth,trim=0 0 0 0,clip]{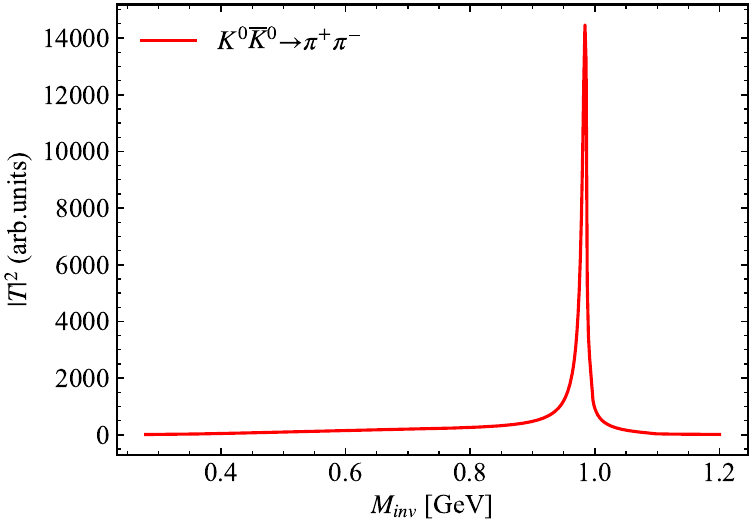} 
\label{fig:T41}
\end{minipage}
\begin{minipage}{0.33\linewidth}
\centering
\includegraphics[width=1\linewidth,trim=0 0 0 0,clip]{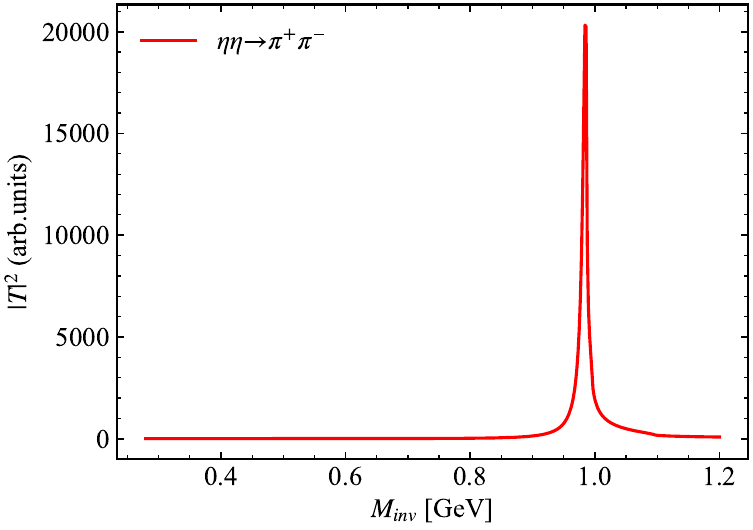} 
\label{fig:T51}
\end{minipage}
\caption{Modulus square of the amplitudes in coupled channels.}
\label{fig:TXX}
\end{figure*}

In the $D^{+}\rightarrow \pi^{+}\pi^{+}\pi^{-}$ decay, the parameters $\mathcal{D}_{1}$ and $\mathcal{D}_{2}$ are correlated with the phase, so we performed a combined fit for the magnitude and phase of the $\pi^{+}\pi^{-}$ $S$-wave amplitude.
The values of the parameters from the fit are listed in Table \ref{tab:DParameter1} and the results are shown in Fig. \ref{fig:DAandP}.
Our results could be a good description of the LHCb measurements for the magnitude and phase of the $\pi^{+}\pi^{-}$ amplitude.
The broad bump structure in the $\pi^{+}\pi^{-}$ amplitude from $\pi\pi$  to $K\bar{K}$ thresholds represents the signal of the dynamical generation of the $f_{0}(500)$ resonance.
Whereas the narrow peak at $0.98$ GeV in our results represents the $f_{0}(980)$ resonance, which shows the cusp effect near the $K\bar{K}$ threshold.
Similar signals also appear in the phase of the $\pi^{+}\pi^{-}$ amplitude.
This is the natural result of Eq. \eqref{eq:DA} which includes all five coupled channels.
The modules of the corresponding coupled channel amplitudes at $\mu=0.6$ GeV are shown in Fig. \ref{fig:TXX}.
The resonance of $f_{0}(980)$ appears in all the coupled channels, while $f_{0}(500)$ appears only in the $\pi^{+}\pi^{-}$ and $\pi^{0}\pi^{0}$ channels.
Furthermore, our theoretical results for the phase of the $D^{+}$ decay are better than those for the $D_{s}^{+}$ decay.

As mentioned above, our calculation of the amplitude of the $D_{s}^{+}$ decay did not fully describe the experimental data, as shown in Fig. \ref{fig:DsAandP}.
From the size of magnitude of the amplitude, we think that this may be due to the small width of the $f_{0}(980)$ resonance in our model.
Therefore, we consider the interactions that do not include the $\eta\eta$ channel, as we found that the width of $f_{0}(980)$ increases in this case in Ref. \cite{Ahmed:2020qkv}.
The refitted parameters without the $\eta\eta$ channel are given in Tables \ref{tab:DsParameter2} and \ref{tab:DParameter2}, where $\mu=0.931$ GeV is taken from Refs. \cite{Xiao:2019lrj,Ahmed:2020kmp,Ahmed:2020qkv} and obtained from the combined fit to several sets of experimental data.

\begin{table}[htbp]
\centering
\renewcommand\tabcolsep{2.0mm}
\renewcommand{\arraystretch}{1.50}
\caption{Values of the parameters from the fit in the $D_{s}^{+}$ decay excluding the $\eta\eta$ channel.}
\begin{tabular*}{86mm}{@{\extracolsep{\fill}}l|cc}
\toprule[1.00pt]
\toprule[1.00pt]
Parameter&$\mu$&$\mathcal{C}$\\
\hline
Fit I&$0.931$ GeV (fixed)&$13.15\pm0.07$\\
\hline
Fit II&$0.894\pm0.005$ GeV&$13.87\pm0.13$\\
\bottomrule[1.00pt]
\bottomrule[1.00pt]
\end{tabular*}
\label{tab:DsParameter2}
\end{table}
\begin{table}[htbp]
\centering
\renewcommand\tabcolsep{2.0mm}
\renewcommand{\arraystretch}{1.50}
\caption{Values of the parameters from the fit in the $D^{+}$ decay excluding the $\eta\eta$ channel.}
\begin{tabular*}{86mm}{@{\extracolsep{\fill}}l|ccc}
\toprule[1.00pt]
\toprule[1.00pt]
Parameter&$\mu$&$\mathcal{D}_{1}$&$\mathcal{D}_{2}$\\
\hline
Fit I&$0.931$ GeV (fixed)&$9.89\pm0.13$&$-8.01\pm0.11$\\
\hline
Fit II &$0.980\pm0.013$ GeV&$10.03\pm0.13$&$-7.50\pm0.16$\\
\bottomrule[1.00pt]
\bottomrule[1.00pt]
\end{tabular*}
\label{tab:DParameter2}
\end{table}

\begin{figure*}[htbp]
\begin{minipage}{0.4\linewidth}
\centering
\includegraphics[width=1\linewidth,trim=0 0 0 0,clip]{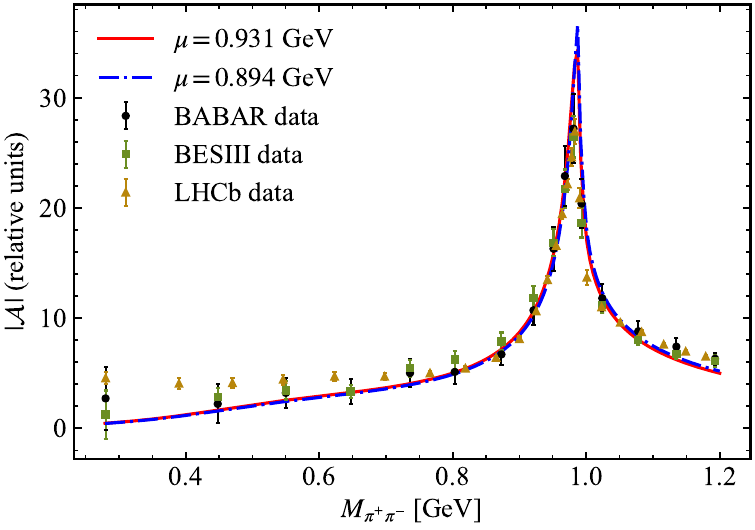} 
\label{fig:ADs2}
\end{minipage}
\quad
\quad
\begin{minipage}{0.4\linewidth}
\centering
\includegraphics[width=1\linewidth,trim=0 0 0 0,clip]{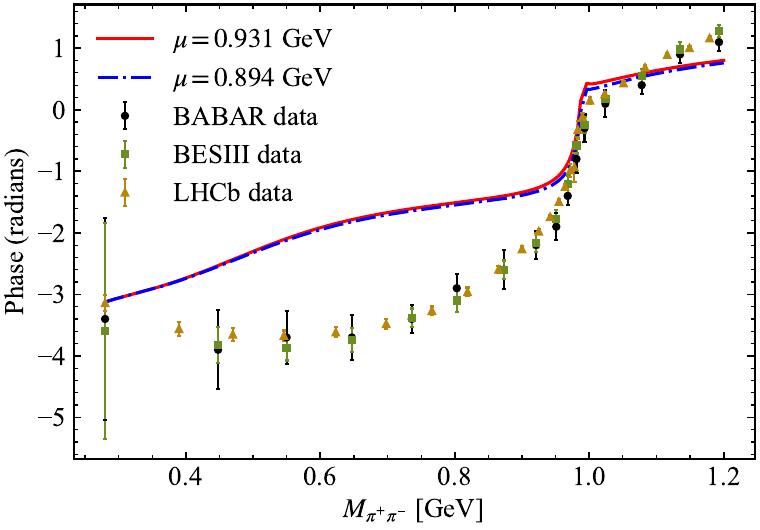} 
\label{fig:PDs2}
\end{minipage}
\begin{minipage}{0.4\linewidth}
\centering
\includegraphics[width=1\linewidth,trim=0 0 0 0,clip]{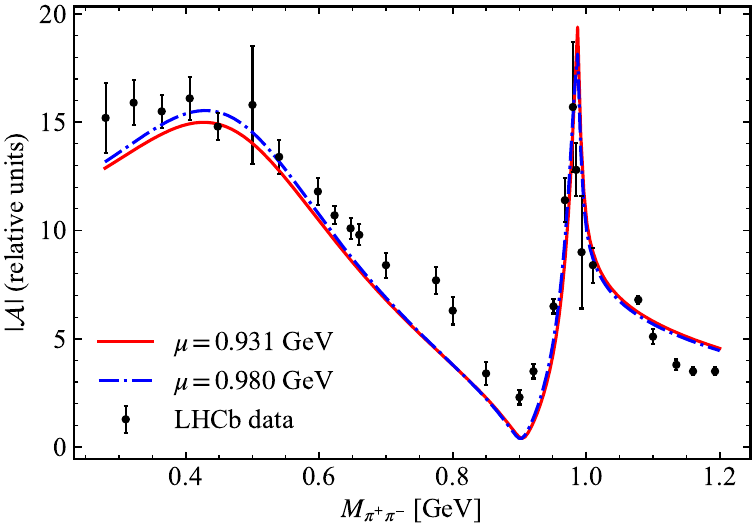} 
\label{fig:ADp2}
\end{minipage}
\quad
\quad
\begin{minipage}{0.4\linewidth}
\centering
\includegraphics[width=1\linewidth,trim=0 0 0 0,clip]{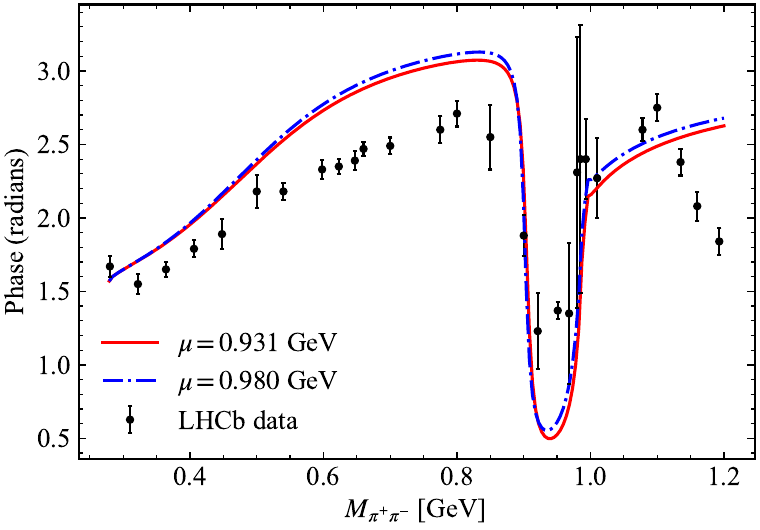} 
\label{fig:PDp2}
\end{minipage}
\caption{The magnitudes (left) and phases (right) of the $\pi^{+}\pi^{-}$ $S$-wave amplitude in the $D_{s}^{+}\rightarrow \pi^{+}\pi^{+}\pi^{-}$ (above row) and $D^{+}\rightarrow \pi^{+}\pi^{+}\pi^{-}$ (below row) decays excluding the $\eta\eta$ channel.}
\label{fig:DsandD}
\end{figure*}

With these fitted parameters in Tables \ref{tab:DsParameter2} and \ref{tab:DParameter2}, we calculate the magnitudes and phases of the $\pi^{+}\pi^{-}$ amplitude in the $D_{s}^{+}\rightarrow \pi^{+}\pi^{+}\pi^{-}$ and $D^{+}\rightarrow \pi^{+}\pi^{+}\pi^{-}$ decays without the $\eta\eta$ channel, which is shown in Fig. \ref{fig:DsandD}.
Our results for the magnitude of the amplitude in the $D_{s}^{+}$ decay describe the experimental data better than including the $\eta\eta$ channel, but there is no significant improvement in the results of the other three figures.

\section{Summary}\label{sec:Summary}

Motivated by the LHCb Collaboration measurements on the processes of $D_{s}^{+}\rightarrow \pi^{+}\pi^{+}\pi^{-}$ and $D^{+}\rightarrow \pi^{+}\pi^{+}\pi^{-}$, we investigate these two decay amplitudes using the chiral unitary approach.
Taking into account the contribution of the $S$-wave interaction between pseudoscalar and pseudoscalar mesons, the $f_{0}(500)$ and $f_{0}(980)$ resonances are dynamically generated as molecular states.
We perform a fit to the magnitude and phase of the $\pi^{+}\pi^{-}$ $S$-wave amplitudes, where our results give a good description for the experimental data.
With the dominant external and internal $W$-emission mechanisms of the $D_{(s)}^{+}$ weak decay, the calculated amplitudes based on the final state interaction can explain why only the $f_{0}(980)$ state appears in the $D_{s}^{+}\rightarrow\pi^{+}\pi^{+}\pi^{-}$ decay, while the $f_{0}(500)$ and $f_{0}(980)$ states appear simultaneously in the $D^{+}\rightarrow\pi^{+}\pi^{+}\pi^{-}$ decay.
This result supports our view that $f_{0}(500)$ is the resonance of $\pi\pi$ and $f_{0}(980)$ is the quasi-bound state of $K\bar{K}$.
In addition, we also consider interactions that do not include the $\eta\eta$ channel, leading to the state of $f_{0}(980)$ with a larger width, and our results can also describe the experimental data well.

\section*{Acknowledgements}

We would like to thank Professor Chu-Wen Xiao for his valuable comments.
This work is supported by the Natural Science Special Research Foundation of Guizhou University Grant No. 2024028 (Z.-Y. W.), and partly by the Fundamental Research Funds for the Central Universities of Central South University under Grants No. 1053320214315, 2022ZZTS0169, and the Postgraduate Scientific Research Innovation Project of Hunan Province under No. CX20220255.

\addcontentsline{toc}{section}{References}

\end{document}